\documentclass[aps,epsf,nopacs,amsmath,amssymb,graphics,preprint]{revtex4}

\usepackage{graphicx}% Include figure files
\usepackage{dcolumn}% Align table columns on decimal point
\usepackage{bm}% bold math
\usepackage{color}
\usepackage{amsmath}

\newcommand{\be}{\begin{equation}}
\newcommand{\ee}{\end{equation}}
\newcommand{\ba}{\begin{eqnarray}}
\newcommand{\ea}{\end{eqnarray}}
\newcommand{\ban}{\begin{eqnarray*}}
\newcommand{\ean}{\end{eqnarray*}}

\newcommand{\eq}[1]{(\ref{#1})}

\newcommand{\A}{\ensuremath{{\cal A}}}
\newcommand{\cL}{\ensuremath{{\cal L}}}
\newcommand{\B}{\ensuremath{{\cal B}}}

\begin{document}
\title{Timescale for trans-Planckian collisions in Kerr spacetime}
\author{
$^{1,2}$Mandar Patil\footnote{Electronic address: mandar@iucaa.ernet.in},
$^3$Pankaj S. Joshi\footnote{Electronic address: psj@tifr.res.in},
$^4$Ken-ichi Nakao\footnote{Electronic address: knakao@sci.osaka-cu.ac.jp},\\
$^5$Masashi Kimura\footnote{Electronic address: M.Kimura@damtp.cam.ac.uk}, 
$^2$Tomohiro Harada\footnote{Electronic address: harada@rikkyo.ac.jp}, 
}

\affiliation{$^1$Inter University Center for Astronomy and Astrophysics Post Bag 4, Ganeshkhind, Pune-411007, India. \\
$^2$Department of Physics, Rikkyo University, Toshima-ku, Tokyo 171-8501 Japan.
\\
$^3$Tata Institute of Fundamental Research, Homi Bhabha Road, Mumbai 400005, India. \\
$^4$Department of Mathematics and Physics, Graduate School of Science, Osaka City University, Osaka 558-8585, Japan. \\
$^5$DAMTP, Centre for Mathematical Sciences, University of Cambridge, Wilberforce Road, Cambridge CB3 0WA, United Kingdom.
}

\date{\today}

\begin{abstract}
We make a critical comparison between
ultra-high energy particle collisions around an extremal Kerr black hole
and that around an over-spinning Kerr singularity,
mainly focusing on the issue of the timescale of collisions.
We show that the time required for 
two massive particles with the proton mass or two massless particles 
of GeV energies to collide around the Kerr black hole with 
Planck energy is several orders of magnitude longer
than the age of the Universe for astro-physically relevant masses 
of black holes, whereas time required  
in the over-spinning case
is of the order of ten million years 
which is much shorter than the age of the Universe.
Thus from the point of view of observation of 
Planck scale collisions, the over-spinning Kerr geometry, 
subject to their occurrence, has distinct advantage over their black hole counterparts.
\end{abstract}

\preprint{OCU-PHYS-421}
\preprint{AP-GR-121}
\preprint{RUP-15-7}

\maketitle

%%%%%%%%%%%%%%%%%%%%%
\section{Introduction}
%%%%%%%%%%%%%%%%%%%%%
Ba\~nados, Silk and West pointed out that 
a Kerr black hole can be a very effective collider for any massive particles~\cite{BSW,BSW2} 
(see also recent review~\cite{Harada:2014vka}), and 
this process can be extended to the massless particles as well~\cite{HK}. 
Around a maximally spinning black hole whose absolute value of the angular momentum $J$ is equal to the 
threshold value $J_{\rm th}:=GM^2/c$, or in other words, an extreme Kerr black hole, 
there is a class of timelike and null geodesics  
that asymptotically approach the black hole from a distant place
 and wind the event horizon infinite times, 
where $M$ is the ADM mass of the black hole, and $c$ and $G$ are the speed of light and Newton's 
gravitational constant, respectively.
If the massive particle has a critical value of the specific angular momentum 
$L=L_{\rm c}:=2GM{\cal E}/c$, where ${\cal E}$ is the 
specific energy or if the impact parameter of the massless particle, which 
is equal to its angular momentum divided by its energy, 
takes a critical
value $B=B_{\rm c}=2GM/c^3$, it may move on such a geodesic. 
If another massive particle with $L\neq L_{\rm c}$ 
or massless particle with $B\neq B_{\rm c}$  starts to fall from a distant place and collides with 
the particle going ahead, the energy of collision at their center of mass frame can be indefinitely large. 
We can recognize the physical mechanism of this phenomena for massive particles as follows~\cite{Jacobson}. 
The event horizon is generated by the outward null geodesic congruence. This fact implies that 
the world line of the particle asymptotically approaching the event horizon becomes asymptotically 
outward null. Hence the relative velocity between the particle with $L=L_{\rm c}$ and the other particle  
with $L\neq L_{\rm c}$ may be arbitrarily close to the speed of light at the collision 
event near the event horizon. This large relative velocity causes 
the large center of mass energy.  Hereafter, we call this phenomenon the BSW process. 
Since this process is theoretically fascinating, many subsequent studies 
appeared~\cite{Berti,Grib-1,Grib-2,HNM,BHaccel}.

A similar possibility for the BSW process but in the 
the over-spinning Kerr spacetime which has $J>J_{\rm th}$, 
has been pointed out by two of present authors, MP and PSJ~\cite{PJ-1,PJ-2}. 
In the case of the over-spinning Kerr spacetime, there is no event horizon and hence there is a class of 
timelike geodesics which approach 
the spacetime singularity
 from a distant place
but turn outward after going through the peri-singularity. 
A particle going outward along such a timelike geodesic can collide with the other ingoing particle.
MP and PSJ showed that the energy of such a collision defined in the 
center of mass frame can be indefinitely large in the limit of $J\rightarrow J_{\rm th}+\epsilon$ 
($0<\epsilon \ll1$), 
if the collision occurs at a special place; $r=GM/c^2$ in the Boyer-Lindquist coordinate for the Kerr spacetime. 
Hereafter we call this phenomenon the PJ process.

We should note that the existence of the spacetime singularity is not 
necessary for the occurrence of particle collisions through PJ process, since
the collision point $r = GM/c^2$ is far from the spacetime
singularity which is located at $r = 0$.
We can consider PJ process in the over-spinning Kerr geometry
temporarily produced by a rapidly rotating regular compact matter.
Such the scenario has been recently suggested in~\cite{NKHPJ}.

Both BSW process and PJ process are theoretically fascinating, but several drawbacks have 
been pointed out from observational point of view. In this paper, we reconsider these drawbacks, especially, 
the timescale issue.

%%%%%%%%%%%%%%%%%%%%%
\section{Drawbacks}
%%%%%%%%%%%%%%%%%%%%%

The following drawbacks of the trans-Planckian BSW process of elementary particles in our Universe 
have been pointed out from the observational point of view;  

\begin{enumerate}
\item The angular momentum $J$ of the black hole is bounded above by 
Thorne's limit $0.998J_{\rm th}$~\cite{Thorne,Berti}. 
\item The effect of the self-gravity may change the trajectories of falling particles and 
will suppress the energy of collision  between particles~\cite{Berti}.
\item The initial conditions of falling particles must be finely tuned so that the collision energy 
exceeds the Planck scale~\cite{Jacobson}. 
\item A too long timescale may be necessary for its occurrence~\cite{Grib-1}. 
\item The flux emitted from BSW process
may become unmeasurably small due to strong redshift and diminished escape fraction~\cite{McWilliams:2012nx}.
\end{enumerate}

On the first point, Abramowicz and Lasota have pointed out 
that Thorne limit $J/J_{\rm th}\lesssim 0.998$ is not strict and the maximal 
value of $J/J_{\rm th}$ depends on the assumed accretion disk model~\cite{AL}; 
a detailed analysis has been given in~\cite{S-etal}. 
This means that the astrophysical upper bound on the Kerr parameter is still unclear. 
It is worthwhile to notice that 
Grib and Pavlov have shown that the high energy collision of particles due to the gravity 
can occur even in the case of non-extreme Kerr black hole $J<J_{\rm th}$~\cite{Grib-1}.  

On the second point, the effect of the dissipative self-gravity, i.e., the gravitational radiation is 
not so important if the rest masses of particles are much smaller than the black hole 
mass~\cite{Vitor, Harada:2011pg}.
By contrast, it is not so easy to see 
whether the non-dissipative self-gravity is important in BSW process. 
A similar process can occur in the system of extremal Reissner-Nordstr\"{o}m (RN) 
black hole: when a radially falling charged particle 
with charge $q$ equal to its mass $m$ times 
$\sqrt{4\pi \varepsilon_0 G}$ is asymptotically 
approaching the event horizon from infinity with vanishing initial velocity, 
if it collides with another radially falling 
neutral particle, the energy of the collision in their center of mass frame can be 
arbitrarily large~\cite{Zasl}, where $\varepsilon_0$ is the conductivity of vacuum.  
Two of the present authors, MK and KN, and their collaborator, H. Tagoshi, 
studied the effect of the self-gravity of the particles in a similar system by studying 
the collision between charged and neutral spherical shells around a RN black hole, 
instead of test particles~\cite{KNT}: The motion of the spherical shell can be seen analytically 
by using Israel's formalism~\cite{Israel}. 
They showed that the non-dissipative self-gravity puts an upper 
bound on the collision energy and
the upper bound cannot be trans-Planckian~\cite{KNT}
for typical astrophysical parameters.
This result suggests that the self-gravity  
limits the collision energy in the BSW process, 
although there is no definite analysis on this issue in the case of a rotating black hole. 
In this study, they also showed that even if there is no black hole,  
the high energy collision can occur; 
only by the self-gravity of the shells, the BSW-like collision 
between the constituent particles of the shells can occur~\cite{KNT}.  

The third and fourth issues, i.e., the fine tuning of the angular momentum of 
particles and timescale problems still remain; these problems exist also in the 
process pointed out by Grib and Pavlov~\cite{Grib-2}. It was shown by two of the 
present authors MK and TH that the fine tuning of the angular momentum can be realised 
naturally for the particles orbiting innermost stable circular orbit and thus the third issue
has been addressed to some extent~\cite{HK2}.

On the fifth point,
McWilliams showed 
that the flux directly emitted from the conventional BSW process~\cite{BSW2}
is unmeasurably small because of strong redshift as well 
as greatly diminished escape fraction~\cite{McWilliams:2012nx}.
However, we should note that 
the potential indirect observability 
was discussed in~\cite{Gariel:2014ara}.
Recently,
the possibilities of high energy debris
by using the efficient energy extractions from a black hole 
were also discussed in~\cite{Schnittman:2014zsa, Berti:2014lva}.
We consider this topic is still challenging problem to be solved.

The same obstacles as those of the BSW process except for the first one seem to 
exist also in the PJ process; the first issue on the upper bound on the angular momentum 
of the black hole is replaced by

\begin{enumerate}
\item The cosmic censorship hypothesis implies that no over-spinning Kerr singularity 
exists in our Universe~\cite{penrose1969,penrose1979}. 
\end{enumerate}

A process similar to PJ one can occur in the over-charged RN spacetime in which the 
spacetime singularity is naked~\cite{PJKN}. Due to the repulsive nature of the central naked singularity, 
a radially falling neutral particle eventually turns to the outward radial direction, and if it collides with 
another radially falling particle at the minimum of its effective potential which corresponds to 
the classical radius $Q^2/4\pi\varepsilon_0c^2M$, the energy of collision in the center of mass frame 
can be arbitrarily large in the limit of $Q/\sqrt{4\pi\varepsilon_0}M\rightarrow1_{+0}$. 
In~\cite{PJKN}, by studying the collision between two spherical dust shells
around an over charged RN spacetime, it is shown that the non-dissipative self-gravity does not 
prevent the trans-Planckian collision unlike the study in~\cite{KNT}.
If one of the two shells is charged, no naked singularity is necessary for 
the high energy collision between the charged shell and 
the other neutral shell by virtue of their self-gravity~\cite{NKPJ}.
This result implies that even in the case with no naked singularity, PJ process 
can occur; we should note that the naked singularity itself is not necessary but the geometry around 
the naked singularity is. Recently, the present authors numerically studied the initial data of a rapidly  
rotating shell whose outside is equivalent to the spacelike hypersurface of the 
over-spinning Kerr spacetime $J>J_{\rm th}$,  
in order to see how small the over-spinning body can be in general relativity; the result implies that 
the over-spinning shell can be so small that the PJ process may occur around it~\cite{NKHPJ}. 
Hence, even if the cosmic censorship conjecture is true, PJ process can occur, in principle.

The fine tuning and timescale issues still exist also in the PJ process, and in this paper, we study 
the timescale for distant observers for which the BSW or PJ process occurs, in more detail. 
In the next section, we investigate the timescale for the occurrences of 
particle collisions through BSW and PJ processes. 

Hereafter, we basically adopt the geometrized unit in which the speed of light 
and the Newton gravitational constant are one. If necessary, we denote the speed 
of light and the Newton's gravitational constant by $c$ and $G$, respectively again.  

%%%%%%%%%%%%%%%%%%%%%
\section{Timescale}
%%%%%%%%%%%%%%%%%%%%%

Consider Kerr spacetime with the mass parameter $M$ and 
the Kerr parameter $a$ which is equivalent to the specific angular momentum of the system. 
The infinitesimal world interval is given by
\begin{equation}
ds^2=-\frac{\varSigma\varDelta}{A}dt^2+\frac{A}{\varSigma}\sin^2\theta\left(d\varphi-\frac{2aMr}{A}dt\right)^2
+\frac{\varSigma}{\varDelta}dr^2+\varSigma d\theta^2,
\end{equation}
where
\begin{eqnarray}
\varDelta&=&r^2-2Mr+a^2, \\
\varSigma&=&r^2+a^2\cos^2\theta, \\
A&=&(r^2+a^2)^2-a^2\varDelta\sin^2\theta.
\end{eqnarray}
For simplicity, we consider the geodesic motion of a particle which is restricted in the equatorial 
plane $\theta=\pi/2$. The radial and time components of its 4-momentum is given by 
\begin{eqnarray}
P^r &=&\pm \frac{1}{r^2}\sqrt{[E(r^2+a^2)-La]^2-\varDelta \left[m^2r^2+(L-aE)^2\right]},
\label{pr} \\
P^t &=&\frac{1}{r^2\varDelta}(r^2+a^2)[E(r^2+a^2)-La]-\frac{a}{r^2}(aE-L),
\label{pt}
\end{eqnarray}
where $E$, $L$ stand for the conserved energy and angular momentum of the particle. 

We consider a collision between two particles with mass $m_I$, 
the conserved energy $E_I$ and the conserved 
angular momentum $L_I$ ($I=1,~2$) at the radial location $r$.  
The energy of collision defined in their center of mass frame 
is given by~\cite{HK} 
\begin{equation}
E_{\rm cm}^2=m_1^2+m_2^2
+\frac{2}{\varSigma}\left[\frac{P_1 P_2 \mp \sqrt{R_1}\sqrt{R_2}}{\varDelta}-\left(L_1-aE_1\right)\left(L_2-aE_2\right)\right]
\end{equation}
where 
\begin{eqnarray}
 P_I(r)&=&\left(r^2+a^2\right)E_I-aL_I \\
 R_I(r)&=& P_I^2(r)-\varDelta \left[m_I^2r^2+\left(L_I-aE_I\right)^2\right]
 \end{eqnarray}
Here plus and minus signs stand for the cases where radial 
velocities of the two particles have opposite and 
same sign, respectively.

%%%%%%%%%%%%%%%%%%%%%%%%%%%%%%%%%%%%%%%%%%%%%%%%
\subsection{Collision between massive particles}
%%%%%%%%%%%%%%%%%%%%%%%%%%%%%%%%%%%%%%%%%%%%%%%%

We first deal with the case where colliding particles are massive.

We introduce following dimensionless variables 
$$
T=\frac{t}{M},~X=\frac{r}{M},~\A=\frac{a}{M},~ {\cal E}=\frac{E}{m},~\cL=\frac{L}{mM}.
$$ 
Then, from Eqs.~\eq{pr} and \eq{pt}, we have
\begin{equation}
 \frac{dX}{dT}=\pm \frac{\sqrt{\left[{\cal E}(X^2+\A^2)-\cL\A\right]^2
 -(X^2-2X+\A^2)\left[X^2+(\cL-\A{\cal E})^2\right]}\left(X^2-2X+\A^2\right)}{(X^2+\A^2)
 \left[{\cal E}(X^2+\A^2)-\cL\A\right]-\A(\A {\cal E}- \cL)\left(X^2-2X+\A^2\right)}
\end{equation}

Hereafter, we restrict ourselves to the marginally bound case ${\cal E}=1$. 
Then, we have
\begin{equation}
\frac{dX}{dT}=\pm\frac{1}{I(X)},
\end{equation}
where
\begin{equation}
I(x;{\cL})=\frac{\sqrt{x}\left[x^3+\A^2x+2\A(\A-\cL)\right]}
{D(x)\sqrt{2x^2-\cL ^2 x+2(\cL-\A)^2}},
\label{I-def}
\end{equation}
with
\begin{equation}
D(x)=x^2-2x+\A^2. \label{D-def}
\end{equation}

We consider the collision between two identical particles of 
mass $m$ in the Kerr spacetime. The energy $E_{\rm cm}$ of the collision in their 
center of mass frame  is given by 
\begin{equation}
\label{ecm}
\frac{E_{\rm cm}^2}{2m^2}=1+\frac{1}{X_{\rm c}^{2}}\left[\frac{\bar{P}_1(X_{\rm c}) 
\bar{P}_2(X_{\rm c}) \mp \sqrt{\bar{R}_1(X_{\rm c})}\sqrt{\bar{R}_2(X_{\rm c})}}
{D(X_{\rm c})}-\left(\cL_1-\A\right)\left(\cL_2-\A\right)\right]
\end{equation}
where
\begin{eqnarray}
 \bar{P}_I(x)&=& x^2+\A^2-\A\cL_I \\
 \bar{R}_I(x)&=& \bar{P}_I^2(x)-D(x) \left[x^2+\left(\cL_I-\A\right)^2\right]
\end{eqnarray}
$X_{\rm c}$ is the dimensionless radial coordinate $X$ 
at the collision event, and $\cL_1$ and $\cL_2$ are  
the dimensionless specific angular momenta of two particles, respectively  

%%%%%%%%%%%%%%%%%%%%%
\subsubsection{The case of the BSW process}
%%%%%%%%%%%%%%%%%%%%%

First of all, let us focus on the collision of two particles around the extremal black hole $\A=1$
close to the event horizon $X=1$. We assume that one of the particle has critical angular
momentum $\cL_1=2$ and other particle has subcritical angular momentum $\cL_2=\cL$ ($<2$).
Since in this case, the radial velocities of two particles have identical sign, 
the energy of collision between these two particles in their center of mass frame is given by
\begin{equation}
 \frac{E_{\rm cm}}{m}\approx \sqrt{\frac{2(2-\sqrt{2})(2-\cL)}{X_{\rm c}-1}}:  
\label{ecm1}
\end{equation}
See Eq.~(2.18) in Ref.~\cite{HNM}.
For the first particle with critical angular momentum, $\cL_1=2$, we have 
\begin{equation}
 I(x;2)=\frac{\sqrt{x}(x^2+x+2)}{\sqrt{2}(x-1)^2}
\end{equation}
In this case, 
the time required 
to reach the collision point $X_{\rm c}$ from a distant location $X_{\rm i}$ is given by 
\begin{equation}
T=-\int^{X_{\rm c}}_{X_{\rm i}}I(x;2)dx 
=f(X_{\rm i})-f(X_{\rm c}), \label{tmbh}
\end{equation}
where
\begin{equation}
 f(x)=\frac{\sqrt{2x}}{3(x-1)}\left(x^2+8x-15\right)+\frac{5}{\sqrt{2}}\ln\left(\frac{\sqrt{x}-1}{\sqrt{x}+1}\right)
\end{equation}

We assume that $X_{\rm i}$ is much larger than unity but much less than $(X_{\rm c}-1)^{-1}$: 
This assumption is valid in reasonable astrophysical situations. 
Then, from Eqs.~\eq{ecm1} and  \eq{tmbh}, we have
\begin{equation}
t=\frac{GM}{c^3}T\simeq
6.5\times10^{25} (2-\cL)^{-1}\left(\frac{M}{M_\odot}\right)\left(\frac{1{\rm GeV}}{mc^2}\right)^2
\left(\frac{E_{\rm cm}}{E_{\rm pl}}\right)^2 {\rm yr},
\end{equation}
where $M_\odot$ and $E_{\rm pl}$ are the solar mass ($1.989\times10^{30}$kg) and the Planck energy ($1.221\times10^{19}$GeV), respectively. The above result implies that much longer time than 
the age of the universe is necessary so that distant observers detect something emitted from 
the collision with the energy comparable to the Planck energy.

%%%%%%%%%%%%%%%%%%%%%
\subsubsection{The case of PJ process}
%%%%%%%%%%%%%%%%%%%%%

We now deal with the Kerr naked singularity case. 
The spin parameter is $\A=1+\epsilon$ with $\epsilon \rightarrow 0^{+}$.
The collision takes place at $X_{\rm c}= 1$ (the minimum of $\varDelta$).
The first particle is initially ingoing but eventually turns back 
at the ``peri-singularity" $X=X_{\rm p}<1$ and arrives at $X_{\rm c}=1$ as an outgoing particle, where 
$$
X_{\rm p}=\frac{1}{4}\left[\cL^2+\sqrt{\cL^4-16(\cL-\A)^2}\right].
$$
It must have angular momentum in the range $2(-1+\sqrt{1+\A})<\cL<(2\A-\sqrt{2\A^2-2})$: 
See Eq.~(19) in Ref.~\cite{PJ-1}. 
The lower limit on the angular momentum is to make sure that the particle indeed turns back and does not hit the singularity as the ingoing particle. 
The upper limit is to make sure that particle turns back at the radial coordinate $x<1$. 
The second particle arrives at the collision point as an ingoing particle
with angular momentum $0<\cL<(2\A-\sqrt{2\A^2-2})$. The upper limit imposed on the angular momentum is to make sure that 
it reaches $x=1$ as an ingoing particle from infinity. We assumed that the angular 
momentum of both the particles is positive which need not be the case. 

Since in this case, the two particles have the radial velocities with signs different from each other, we 
have, from Eq.~(\ref{ecm}) with the positive sign, the collision energy $E_{\rm cm}$ at their 
center of mass frame which is given by 
\begin{equation}
 \frac{E_{\rm cm}}{m}\approx \frac{2}{\sqrt{\epsilon}}\sqrt{(2-\cL_1)(2-\cL_2)}: 
\label{ecm2}
\end{equation}
See Eq.~(20) in Ref.~\cite{PJ-1}. 

Let us estimate the time required for the high energy collision of particles. 
The time required for the 
first particle to reach the collision point $X_{\rm c}=1$ 
after going through the peri-singularity at $X=X_{\rm p}$ is given by 
\begin{equation}
T=-\int_{X_{\rm i}}^{X_{\rm p}}I(x;\cL_1)dx+\int_{X_{\rm p}}^1 I(x;\cL_1)dx. \label{T-integral-2}
\end{equation}
Here note that $T$ is dominated by the integral in the neighborhood of $x=1$ because 
$0<D(1)=\A^2-1=\epsilon(2+\epsilon) \ll1$. In the limit of $\epsilon\rightarrow 0$, $T$ becomes 
infinite. 
In order to estimate the integral in \eq{T-integral-2}, we make use of the standard result for the 
Lorentzian distribution: For $0<\A^2-1\ll1$, 
\begin{equation}
\frac{1}{D(x)}=\frac{\pi}{\sqrt{\A^2-1}}\left[\frac{1}{\pi}\frac{\sqrt{\A^2-1}}{(x-1)^2+\A^2-1}\right]
\simeq \frac{\pi}{\sqrt{\A^2-1}} \delta(x-1)
\label{loren}
\end{equation}
The time required for the first particle to reach the collision point $X=1$ 
after going through the peri-singularity at $X=X_{\rm p}$ is estimated as $T\simeq 
3\pi/\sqrt{2\epsilon}$, and by using Eq.~\eq{ecm2}, we have
\begin{equation}
t=\frac{GM}{c^3}T\simeq 6.9\times10^6 \left(2-\cL_1\right)^{-{1\over2}}\left(2-\cL_2\right)^{-{1\over2}}
\left(\frac{M}{M_\odot}\right)\left(\frac{1{\rm GeV}}{mc^2}\right)
\left(\frac{E_{\rm cm}}{E_{\rm pl}}\right){\rm yr}.
\end{equation}
The time necessary for the occurrence of the Planck scale collision between two protons 
can be much less than the age of the Universe. 

%%%%%%%%%%%%%%%%%%%%%%%%%%%%%%%%%%%%%%%%%%%%%%%%
\subsection{Collision between massless particles}
%%%%%%%%%%%%%%%%%%%%%%%%%%%%%%%%%%%%%%%%%%%%%%%%

We now deal with the case where the colliding particles are massless.

Here we introduce one more dimensionless variable, 
$$
\B=\frac{L}{ME}.
$$ 
Then, from Eqs.~\eq{pr} and \eq{pt}, we have
\begin{equation}
 \frac{dX}{dT}=\pm\frac{1}{J(X;\B)}, 
\end{equation}
where
\begin{equation}
J(x;\B)=\frac{\sqrt{x}\left[ x^3+\A^2 x+2\A(\A-\B) \right]}{D(x)\sqrt{x^3+(\A^2-\B^2)x+2(\A-\B)^{2}}},
\end{equation}
with $D(x)$ given by Eq.~(\ref{D-def}).

We consider the collision between two massless particles each with conserved energy $E$. 
The energy $E_{\rm cm}$ of the collision in their 
center of mass frame  is given as
\begin{equation}
\frac{E_{\rm cm}^2}{E^2}=\frac{2}{X_{\rm c}^{2}}\left[\frac{\bar{P}_1(X_{\rm c}) \bar{P}_2(X_{\rm c}) \mp \sqrt{\bar{R}_1(X_{\rm c})}\sqrt{\bar{R}_2(X_{\rm c})}}{D(X_{\rm c})}-\left(\B_1-\A\right)\left(\B_2-\A\right)\right]
\end{equation}
where and $\B_1$ and $\B_2$ are  
the dimensionless impact parameters of the two particles, and
\begin{eqnarray}
 \bar{P}_I(x)&=&x^2+\A^2-\A\B_I, \\
 \bar{R}_I(x)&=& \bar{P}_I^2(x)-D(x)\left(\A-\B_I\right)^2.
\end{eqnarray}

%%%%%%%%%%%%%%%%%%%%%
\subsubsection{The case of BSW process}
%%%%%%%%%%%%%%%%%%%%%

We consider collision of two massless particles around the extremal black hole $\A=1$
near the event horizon $X=1$. We assume that one of the particle has a critical value of the impact parameter
$\B_1=2$ and other particle has subcritical impact parameter $\B_2=\B$ ($<2$).
Again since the radial velocities of the two particles have the identical sign, 
the energy of collision between these two particles in their center of mass frame turns out to be
\begin{equation}
 \frac{E_{\rm cm}}{E}\approx \sqrt{\frac{2(2-\sqrt{3})(2-\B)}{X_{\rm c}-1}}: 
\label{ecm11}
\end{equation}
See Eq.~(2.18) in Ref.~\cite{HNM}.

For the first particle with critical impact parameter $\B_1=2$, we have 
\begin{equation}
 J(x;2)=\frac{\sqrt{x}(x^2+x+2)}{\sqrt{x+2}(x-1)^2}
\end{equation}
The integral can be carried out exactly and we get the following result for the timescale.
\begin{equation}
T=-\int^{X_{\rm c}}_{X_{\rm i}}J(x;2)dx=g(X_{\rm i})-g(X_{\rm c}) \label{tmbh2}
\end{equation}
where
\begin{eqnarray}
 g(x) =
\frac{1}{3}\sqrt{x}\sqrt{x+2}\frac{(3x-7)}{(x-1)}+4~{\rm arcsinh}\left( \frac{\sqrt{x}}{\sqrt{2}}\right)
 +
\frac{13}{3\sqrt{3}} \ln \left[
\frac{2 x-\sqrt{3} \sqrt{x (x+2)}+1}{x-1}
\right].
\end{eqnarray}

It is assumed that $X_{\rm i}$ is much larger than unity but much less than $(X_{\rm c}-1)^{-1}$: 
Then, from Eqs.~\eq{ecm11} and  \eq{tmbh2}, we have
\begin{equation}
t=\frac{GM}{c^3}T\simeq
1.0\times10^{26}
(2-\B)^{-1}\left(\frac{M}{M_\odot}\right)\left(\frac{1{\rm GeV}}{E}\right)^2
\left(\frac{E_{\rm cm}}{E_{\rm pl}}\right)^2 {\rm yr}.
\end{equation}
This result implies that the time required for massless particles with energy 1GeV 
as measured at infinity to reach the collision point
close to the event horizon of the extremal black-hole and participate in 
Planck scale collision is much larger than the age of the universe.
This is similar to the result obtained in the case of massive particles.

%%%%%%%%%%%%%%%%%%%%%
\subsubsection{The case of the PJ process}
%%%%%%%%%%%%%%%%%%%%%

We now consider the 
over-spinning Kerr case with spin parameter ${\cal A}$ slightly larger than unity, i.e., 
$\A=1+\epsilon$ with $\epsilon \rightarrow 0^{+}$. As in the case of massive particles 
the collision between two massless particles takes place at $X_{\rm c}= 1$.

The first particle is initially ingoing and eventually turns back 
at the ``peri-singularity" $X=X_{\rm p}<1$ and arrives at $X_{\rm c}=1$ as an outgoing particle.
It must have dimensionless impact parameter in the range 
\begin{equation}
\A<\B_1<2\A-\sqrt{A^2-1}:
\end{equation} 
See Eq.~(5) and Fig.~2 in Ref.~\cite{MPEF}. As in the case of massive particle, 
the lower limit on the impact parameter $\B_1$ is to make sure that the particle indeed turns back and does not hit the singularity 
as the ingoing particle. The upper limit is to make sure that particle turns back at the radial coordinate $x<1$. 
The second particle is an ingoing particle at $X=1$
with impact parameter $\B_2$ which should satisfy 
\begin{equation}
-\A-\frac{3+3({\cal A}-\sqrt{{\cal A}^2-1})^{2\over3}}{({\cal A}-\sqrt{{\cal A}^2-1})^{1\over3}}
<\B_2<2\A-\sqrt{A^2-1}:
\end{equation}
See Eq.~(5) and Fig.~2 in Ref.~\cite{MPEF}. 
Both the upper and lower limits imposed on the impact parameter $\B_2$ are to make sure that 
it reaches $x=1$ as an ingoing particle from infinity.

Two particles have radial velocities with opposite signs and thus the center of mass energy is given by 
\begin{equation}
 \frac{E_{cm}}{E}\approx \frac{1}{\sqrt{\epsilon}}\sqrt{2(2-\B_1)(2-\B_2)}. 
\label{ecm22}
\end{equation}

The time required for the first particle  
to reach the collision point $X_{\rm c}=1$ 
after going through the peri-singularity at $X=X_{\rm p}$ is given by 
\begin{equation}
T=-\int_{X_{\rm i}}^{X_{\rm p}}J(x;\B_1)dx+\int_{X_{\rm p}}^1 J(x;\B_1)dx. \label{T-integral-3}
\end{equation}
Again using the fact that integral gets dominant contribution from the region close to $X=1$ and 
the standard result for Lorentzian distribution we compute the time required for the 
first particle to reach a collision point after having gone through peri-singularity. We get

\begin{equation}
t=\frac{GM}{c^3}T
\simeq 
9.0\times10^6 ~(2-\B_1)^{-{1\over2}}(2-\B_2)^{-{1\over2}}
\left(\frac{M}{M_\odot}\right)\left(\frac{1{\rm GeV}}{E}\right)
\left(\frac{E_{\rm cm}}{E_{\rm pl}}\right){\rm yr}.
\end{equation}
The time necessary for the occurrence of the Planck scale collision between two 
massless particles of conserved energy 1GeV can be much less than Hubble time.

%%%%%%%%%%%%%%%%%%%%%
\section{Summary and Discussion}
%%%%%%%%%%%%%%%%%%%%%

The time required for the massive particles with $m\approx 1$GeV or 
massless particles with the same energy as $m$ to participate in the Planck scale 
collision $E_{\rm cm}\approx 10^{19}$ GeV around an extremal Kerr black hole 
of mass $M=M_{\odot}$ is about $10^{15}$ times longer than the age 
of the Universe, whereas it can be of no more than of the order of
ten million years in the case of the near-extremal over-spinning Kerr geometry with 
the same mass. 
Here it is worthwhile to notice that the massless particles can be photons 
since the high-energy collision between photons is possible 
through quantum effects. 
We note that the timescale of the BSW like and PJ like processes in the Reissner-Nordstr\"om spacetime
was investigated in~\cite{PJKN}, and the results are the same as in the present study.

The origin of this large difference in the timescale 
is the sign of the radial velocities at the collision event. 
In  the BSW case, both particles are ingoing at the collision event. 
By contrast, in the PJ case, one particle is outgoing but the other is ingoing, at the collision event. 
In  the extreme limit $\A\rightarrow1+0$, the ratio of the timescale of PJ process to 
that of the BSW one vanishes.  
Hence, from the observational point of view, PJ process is more important than BSW process 
subject to the emergence of the over-spinning Kerr geometry in the Universe.

\section*{Acknowledgments}
KN and MK are grateful to H. Ishihara and colleagues in the group of elementary 
particle physics and gravity at Osaka City University for useful discussions.  
K.N was supported in part
by JSPS Grant-in-Aid for Scientifc Research (C) (No. 25400265). 
M.K was partially supported by  JSPS Fellows 
(No.23$\cdot$2182) and grant for research abroad from JSPS. 
T.H. was partially supported by the Grant-in-Aid No. 26400282 for Scientific Research
Fund of the Ministry of Education, Culture, Sports, Science and Technology, Japan.

\end{document}